\documentclass[english]{article}
\usepackage[T1]{fontenc}
\usepackage[latin1]{inputenc}
\usepackage{color}
\usepackage{graphicx}

\makeatletter

\AtBeginDocument{
  
}

\usepackage{babel}
\makeatother
\begin{document}
\begin{center}\textbf{\textcolor{black}{\Large Generation of higher
order nonclassical states via interaction of intense
electromagnetic
field with third order nonlinear medium}}\textcolor{black}{\Large }\\
 \textcolor{black}{\Large }\end{center}{\Large \par}

\begin{center}\textcolor{black}{Anirban Pathak}%
\footnote{\textcolor{black}{email: anirban.pathak@jiit.ac.in}

\textcolor{black}{~~~~~~~~~~~anirbanpathak@yahoo.co.in}%
}\end{center}

\begin{center}\textcolor{black}{Department of Physics, JIIT, A-10,
Sectror-62, Noida, UP-201307, India.}\\
\end{center}

\begin{abstract}
\textcolor{black}{\normalsize Interaction of intense laser beam
with an inversion symmetric third order nonlinear medium is
modeled as a quartic anharmonic oscillator. A first order operator
solution of the model Hamiltonian is used to study the
possibilities of generation of higher order nonclassical states.
It is found that the higher order squeezed and higher order
antibunched states can be produced by this interaction. It is also
shown that the higher order nonclassical states may appear
separately, i.e. a higher order antibunched state is not
essentially higher order squeezed state and vice versa.
}{\normalsize \par}
\end{abstract}

\section{\textcolor{black}{\normalsize Introduction}}

\textcolor{black}{A nonclassical state of the electromagnetic
field is one for which the Glauber-Sudarshan P-function
{[}\ref{elements of quantum optics}{]} is not as well defined as
the probability density is {[}\ref{the:Dodonov-V-V}{]}. To be
precise, if P-function becomes either negative or more singular
than delta function, then we obtain a nonclassical state. The
nonclassical states do not have any classical analogue. Commonly,
standard deviation of an observable is considered to be the most
natural measure of quantum fluctuation {[}\ref{the:orlowski}{]}
associated with that observable and the reduction of quantum
fluctuation below the coherent state level corresponds to a
nonclassical state. For example, an electromagnetic field is said
to be electrically squeezed field if uncertainties in the
quadrature phase observable $X$ reduces below the coherent state
level (i.e. $\left(\Delta X\right)^{2}<\frac{1}{2}$) and
antibunching is defined as a phenomenon in which the fluctuations
in photon number reduces below the Poisson level (i.e.
$\left(\Delta N\right)^{2}<\langle N\rangle$)
{[}\ref{nonclassical}, \ref{hbt}{]}. Standard deviations can also
be combined to form some complex measures of nonclassicality,
which may increase with the increasing nonclassicality. As an
example, we can note that the total noise} of a quantum state
\textcolor{black}{which,} is a measure of the total fluctuations
of the amplitude\textcolor{black}{, increases with the increasing
nonclassicality in the system {[}\ref{the:hilery2}{]}. }

\textcolor{black}{Probably, antibunching and squeezing are the
most popular examples of nonclassical states and people have shown
serious interest on these states since 1960s. But higher order
extension of these nonclassical states are only introduced in late
1980s {[}\ref{hong1}-\ref{lee2}{]}. Among these higher order
nonclassical effects higher order squeezing is studied in detail
{[}\ref{hong1}, \ref{hong2}, \ref{hillery}, \ref{giri}{]} but the
higher order antibunching {[}\ref{lee1}{]} is not yet studied
rigorously. As a result, still we do not have answers to certain
fundamental questions like: Whether higher order antibunching and
higher order squeezing appears simultaneously or not? The present
work aims to provide answers to this question. In order to do so,
in section 2, we have modeled the interaction of an intense laser
beam with an inversion symmetric third order nonlinear medium as a
quartic anharmonic oscillator. A first order operator solution of
the model Hamiltonian is also used to provide time evolution of
some useful operators. In section 3 and 4 we have theoretically
studied the possibilities of generation of the higher order
squeezed and higher order antibunched states respectively. We have
shown that the generation of higher order squeezed and higher
order antibunched states is possible but they may or may not
appear simultaneously. We finish with some comments and concluding
remarks in section 5. }

\section{\textcolor{black}{\normalsize The model: an intense laser beam interacts
with a 3rd order nonlinear medium}}

\textcolor{black}{An intense electromagnetic field interacting
with a dielectric medium induces a macroscopic polarization
($\overrightarrow{P}$) having a general form \begin{equation}
\overrightarrow{P}=\chi_{1}\overrightarrow{E}+\chi_{2}\overrightarrow{E}\overrightarrow{E}+\chi_{3}\overrightarrow{E}\overrightarrow{E}\overrightarrow{E}+....\label{new1}\end{equation}
 where $\overrightarrow{E}$ is the electric field and $\chi_{i}$
is the $i-th$ order susceptibility. Corresponding electromagnetic
energy density is given by \begin{equation}
H_{em}=\frac{1}{8\pi}\left[(\overrightarrow{E}+4\pi\overrightarrow{P}).\overrightarrow{E}+\overrightarrow{B}.\overrightarrow{B}\right]\label{new2}\end{equation}
 where $\overrightarrow{B}$ is the magnetic field. Now, if we consider
an inversion symmetric medium then even order susceptibilities
($\chi_{2}$$,\chi_{4}$ etc.) would vanish. Hence the leading
contribution to the nonlinear polarization in an inversion
symmetric medium comes through the third order susceptibility
($\chi_{3}$). If we neglect the macroscopic magnetization (if any)
then the interaction energy will be proportional to the $4$-th
power of the electric field. Normal mode expansion of the field
ensures that the electric field operator $E_{x}$ for $x$-th mode
is proportional to $(a_{x}+a_{x}^{\dagger})$ and the free field
Hamiltonian is \begin{equation}
H_{0}=\sum_{x}\omega_{x}(a_{x}^{\dagger}a_{x}+\frac{1}{2})\label{n1}\end{equation}
 where we have chosen $\hbar=1$. }

\textcolor{black}{Thus the total Hamiltonian of a physical system
in which a single mode of intense electromagnetic field having
unit frequency interacts with a 3rd order nonlinear non-absorbing
medium is\begin{equation}
\begin{array}{lcl}
H & = & (a^{\dagger}a+\frac{1}{2})+\frac{\lambda}{16}(a^{\dagger}+a)^{4}\\
 & = & \frac{X^{2}}{2}+\frac{\dot{X}^{2}}{2}+\frac{\lambda}{4}X^{4}\end{array}\label{ten.1}\end{equation}
 with\begin{equation}
X=\frac{1}{\sqrt{2}}(a^{\dagger}+a)\label{11}\end{equation}
 and \begin{equation}
\dot{X}=-\frac{i}{\sqrt{2}}(a^{\dagger}-a).\label{eq:11.1}\end{equation}
The parameter $\lambda$ is the coupling constant}
\textcolor{black}{\emph{}}\textcolor{black}{and is a function of
$\chi_{3}$. Here we can note that the silica crystals which are
used to construct optical fibers are example of third order
nonlinear medium. So third order nonlinear medium described by the
Hamiltonian (\ref{ten.1}) is also important from the point of view
of applicability.}

\textcolor{black}{The above Hamiltonian (\ref{ten.1}) represents a
quartic anharmonic oscillator of unit mass and unit frequency. The
equation of motion corresponding to (\ref{ten.1}) is
\begin{equation} \ddot{X}+X+\lambda
X^{3}=0\label{eqm}\end{equation}
 which can not be solved exactly. But in the interaction picture the
potential $V_{I}$ and the time evolution operator $U_{I}$
corresponding to (\ref{ten.1}) are respectively \begin{equation}
V_{I}(t)=\exp(ia^{\dagger}at)\lambda(a+a^{\dagger})^{4}\exp(-ia^{\dagger}at)=\lambda(a\exp(-it)+a^{\dagger}\exp(it))^{4}\label{eq:VI(t)}\end{equation}
and \begin{equation}
U_{I}(t)=1-i\int_{0}^{t}V_{I}(t_{1})dt_{1}+(-i)^{2}\int_{0}^{t}V_{I}(t_{1})dt_{1}\int_{0}^{t_{1}}V_{I}(t_{2})dt_{2}+.....\,\,\,\,.\label{eq:Ui(t)}\end{equation}
Now if we assume \begin{equation}
\int_{0}^{t}V_{I}(t_{1})dt_{1}\int_{0}^{t_{1}}V_{I}(t_{2})dt_{2}\ll1\label{eq:condition}\end{equation}
then we can neglect higher order terms and ( \ref{eq:Ui(t)})
reduces to \begin{equation}
U_{I}(t)=1-i\int_{0}^{t}V_{I}(t_{1})dt_{1}=1-i\int_{0}^{t}\lambda(a\exp(-it_{1})+a^{\dagger}\exp(it_{1}))^{4}dt_{1}\label{eq:uit1}\end{equation}
Thus the first order expression for time evolution of annihilation
operator is }

\textcolor{black}{\begin{equation}
\begin{array}{lcl}
a_{I}(t)=U_{I}^{\dagger}(t)a(0)U_{I}^{\dagger}(t) & = & a-\frac{i\lambda}{8}\left[6ta+6ta^{\dagger}a^{2}+6\exp(it)\sin ta^{\dagger2}a-\exp(2it)\sin(2t)a^{\dagger3}\right.\\
 & + & \left.6\exp(it)\sin ta^{\dagger}+2\exp(-it)\sin ta^{3}\right].\end{array}\label{ev6}\end{equation}
 The derivation of this first order expression (for a more generalized
Hamiltonian) is shown in detail in {[}\ref{fernandez},
\ref{pathak1}{]}. Here we would like to note that the annihilation
operator in the Heisenberg picture $a_{H}$ and that in the
interaction picture $a_{I}$ are related by
$a_{H}=\exp(-it)a_{I}(t)$ and our solution (\ref{ev6}) is valid
only when (\ref{eq:condition}) is satisfied. Physically this
condition implies that the anharmonic term present in
(\ref{ten.1}) provides only a small perturbation. This assumption
is justified because in a third order nonlinear medium the
anharmonic constant $\lambda$, which is a function of $\chi_{3}$,
is very small (i.e $\lambda\ll1)$ {[}\ref{pathak1}{]}.}

\section{\textcolor{black}{\normalsize Higher order squeezing}}

\textcolor{black}{Higher order squeezing is defined in various
ways. The definition which we have used in this work is by Hillery
{[}\ref{hillery}{]}. This definition is different from that of
Hong and Mandel {[}\ref{hong1}{]}. Present definition is also
called amplitude squared squeezing. According to this definition
of higher order squeezing, higher order quadrature variables are
defined as \begin{equation}
Y_{1}=\frac{1}{\sqrt{2}}(a^{\dagger2}+a^{2})\label{eq:highersq1}\end{equation}
 and \begin{equation}
Y_{2}=\frac{i}{\sqrt{2}}(a^{\dagger2}-a^{2})\label{eq:highersq2}\end{equation}
From the commutation relation $[Y_{1'}Y_{2}]=i(4N+2)$ it is easy
to conclude that a state is squeezed in $Y_{1}$ variable if
\begin{equation} (\bigtriangleup\, Y_{1})^{2}<\left\langle
2N+1\right\rangle \label{eq:cond}\end{equation} or if,
\begin{equation} f=(\bigtriangleup\, Y_{1})^{2}-\left\langle
2N+1\right\rangle <0\label{eq:cond2}\end{equation}
 A strenuous but straight forward operator algebra yields} \begin{equation}
\begin{array}{lcl}
\Delta Y_{1}^{2}=\langle Y_{1}^{2}\rangle-\langle Y_{1}\rangle^{2} & = & \left[2|\alpha|^{2}+1-\frac{\lambda}{4}\left[4|\alpha|^{2}\left(2|\alpha|^{2}+3\right)\sin t\sin(2\theta-t)-12|\alpha|^{4}t\sin(4\theta)\right.\right.\\
 & + & 3\left(2|\alpha|^{4}+4|\alpha|^{2}+1\right)\sin^{2}2t-12|\alpha|^{2}\left(2|\alpha|^{2}+3\right)\sin t\sin(2\theta+t)\\
 & - & \left.\left.2|\alpha|^{4}\sin2t\sin\left(2(2\theta-2t)\right)\right]\right]\end{array}\label{eq:dely1}\end{equation}
 \textcolor{black}{where $\alpha=|\alpha|\exp(i\theta)$ is used. Here we would like to note that since we are working
in interaction picture we have to take all the expectation values
with respect to the initial coherent state $|\alpha\rangle$ which
is defined as $a|\alpha\rangle=\alpha|\alpha\rangle$. By taking
the expectation value of $N(t)=a^{\dagger(t)}a(t)$ we obtain
\begin{equation} \langle
N(t)\rangle=|\alpha|^{2}-\frac{\lambda}{4}\left[2|\alpha|^{2}\left(2|\alpha|^{2}+3\right)\sin
t\sin(2\theta-t)-|\alpha|^{4}\sin2t\sin\left(2(2\theta-2t)\right)\right].\label{eq:<n(t)>}\end{equation}
Substituting (\ref{eq:dely1}) and (\ref{eq:<n(t)>}) in
(\ref{eq:cond2}) we obtain a closed form analytic expression for
$f$ as \begin{equation}
\begin{array}{lcl}
f & = & -\frac{3\lambda}{4}\left[-4|\alpha|^{2}(2|\alpha|^{2}+3)\sin t\sin(t+2\theta)-4|\alpha|^{4}t\sin(4\theta)\right.\\
 & + & \left.(2|\alpha|^{4}+4|\alpha|^{2}+1)\sin^{2}(2t)\right]\end{array}\label{eq:result}\end{equation}
} From (\ref{eq:result}) we can observe that $f$ oscillates
between positive and negative values depending upon the phase of
the input coherent light $\theta$ and the interaction time $t$.
Both of these parameters can be tuned to produce higher order
squeezed state and to increase the depth of noclassicality by
increasing the negativity of $f$. If we consider
$\theta=\frac{\pi}{2}$ then \textcolor{black}{(\ref{eq:result})
reduces to \begin{equation}
f=-\frac{3\lambda}{4}\left[4|\alpha|^{2}(2|\alpha|^{2}+3)\sin^{2}t+(2|\alpha|^{4}+4|\alpha|^{2}+1)\sin^{2}(2t)\right]\label{eq:result1}\end{equation}
which is always negative and thus we always have higher order
squeezing. }

\section{\textcolor{black}{\normalsize Higher order antibunching of photons}}

\textcolor{black}{As we have already mentioned the higher order
antibunching {[}\ref{lee1}{]} is not yet studied rigorously. Using
the negativity of P function {[}\ref{elements of quantum
optics}{]}, Lee introduced the criterion for HOA as }

\textcolor{black}{\begin{equation} R(l,m)=\frac{\left\langle
N_{x}^{(l+1)}\right\rangle \left\langle N_{x}^{(m-1)}\right\rangle
}{\left\langle N_{x}^{(l)}\right\rangle \left\langle
N_{x}^{(m)}\right\rangle }-1<0,\label{eq:ho3}\end{equation} where
$N$ is the usual number operator, $N^{(i)}=N(N-1)...(N-i+1)$ is
the $ith$ factorial moment of number operator, $\left\langle
\right\rangle $ denotes the quantum average, $l$ and $m$ are
integers satisfying the conditions $l\leq m\leq1$ and the
subscript $x$ denotes a particular mode. Ba An {[}\ref{ba an}{]}
choose $m=1$ and reduced the criterion of $l$th order antibunching
to \begin{equation} A_{x,l}=\frac{\left\langle
N_{x}^{(l+1)}\right\rangle }{\left\langle N_{x}^{(l)}\right\rangle
\left\langle N_{x}\right\rangle
}-1<0\label{eq:bhuta1}\end{equation} or, \begin{equation}
\left\langle N_{x}^{(l+1)}\right\rangle <\left\langle
N_{x}^{(l)}\right\rangle \left\langle N_{x}\right\rangle
.\label{eq:ba an (cond)}\end{equation} Physically, a state which
is antibunched in $l$th order has to be antibunched in $(l-1)th$
order. Therefore, we can further simplify (\ref{eq:ba an (cond)})
as \begin{equation} \left\langle N_{x}^{(l+1)}\right\rangle
<\left\langle N_{x}^{(l)}\right\rangle \left\langle
N_{x}\right\rangle <\left\langle N_{x}^{(l-1)}\right\rangle
\left\langle N_{x}\right\rangle ^{2}<\left\langle
N_{x}^{(l-2)}\right\rangle \left\langle N_{x}\right\rangle
^{3}<...<\left\langle N_{x}\right\rangle
^{l+1}\label{eq:ineq}\end{equation} and obtain the condition for
$l-th$ order antibunching as \begin{equation} d(l)=\left\langle
N_{x}^{(l+1)}\right\rangle -\left\langle N_{x}\right\rangle
^{l+1}<0.\label{eq:ho21}\end{equation} This simplified criterion
(\ref{eq:ho21}) coincides exactly with the physical criterion of
HOA introduced by Pathak and Garica {[}\ref{the:martin}{]}. }

\textcolor{black}{By using (\ref{ev6}) and the condition for
higher order antibunching (\ref{eq:ho21}), it is easy to derive
that for a third order non-linear medium having inversion
symmetry, we have
\begin{equation}
d(1)=\frac{3\lambda|\alpha|^{2}}{4}\left[2\left(2|\alpha|^{2}+1\right)\sin(t-2\theta)\sin(t)+|\alpha|^{2}\sin\left(2(t-2\theta)\right)\sin(2t)\right],\label{app1}\end{equation}
\begin{equation}
d(2)=\frac{3\lambda|\alpha|^{4}}{2}\left[\sin(t-2\theta)\sin(t)+\sin\left(2(t-2\theta)\right)\sin(2t)\right]\label{app2}\end{equation}
and\begin{equation}
d(3)=\frac{3\lambda|\alpha|^{4}}{4}\sin\left(2(t-2\theta)\right)\sin(2t).\label{app3}\end{equation}
 Antibunching of fourth or higher order can not be observed with a
first order solution of the model Hamiltonian. In order to study
the possibilities of their occurrence we have to use higher order
operator solutions of (\ref{eqm}). }

\textcolor{black}{Equations (\ref{app1}-\ref{app3}) coincides
exactly with our recent result {[}\ref{the:martin}{]} which was
reported as a special case of a generalized Hamiltonian. Here we
can observe that if we choose interaction time $t=2\theta$ then
$d=0$. Therefore, we can observe higher order coherence. But for
$\theta=0$ or $\theta=n\pi$, (i.e. when input is real) $d$ is a
sum of square terms only. So $d$ is always positive and we have
higher order bunching of photons. For other values of phase
$(\theta)$ of input radiation field, value of $d$ oscillates from
positive to negative, so we can observe higher order bunching,
anti-bunching or coherence in the output depending upon the
interaction time $t$.}

\section{\textcolor{black}{\normalsize Summary and concluding remarks}}

\textcolor{black}{From (\ref{eq:result1}) we know that for}
$\theta=\frac{\pi}{2}$ we always obtain higher order squeezed
state. But for $\theta=\frac{\pi}{2}$, (\ref{app2}) and
(\ref{app3}) reduces to \textcolor{black}{\begin{equation}
d(2)=\frac{3\lambda|\alpha|^{4}}{2}\left[-\sin^{2}(t)+\sin^{2}(2t)\right]\label{app2.1}\end{equation}
and\begin{equation}
d(3)=\frac{3\lambda|\alpha|^{4}}{4}\sin^{2}(2t).\label{app3.1}\end{equation}
}respectively. Here we can see that $d(3)$ is always positive for
this particular choice of $\theta$, and therefore, third order
antibunching will not appear simultaneously with the amplitude
squared squeezing. On the other hand $d(2)$ oscillates between
positive (higher order bunching) and negative (higher order
antibunching) values. Thus we can conclude that neither second and
third order antibunching appears simultaneously nor the higher
order squeezing appears simultaneously with the higher order
antibunching. Alternatively, we can state that, in general higher
order nonclassical effects may appear separately. The possibility
of their appearance may be tuned by tuning the phase ($\theta$) of
the input coherent state and interaction time $(t)$.

\textbf{Acknowledgement:} Author is thankful to his student Mr.
Prakash Gupta for his interest in this work and help in correcting
the manuscript.


\begin{thebibliography}{10}
\bibitem{key-18}\textcolor{black}{\label{elements of quantum optics} Meystre P and
Sargent M III, Elements of Quantum Optics, second edition
Springer-Verlag, Berlin 1991. }
\bibitem{key-22}\textcolor{black}{\label{the:Dodonov-V-V}Dodonov V V} and Mon'ko
V I (editors), Theory of Nonclasical states of light, Taylor \&
Francis, New York 2003.
\bibitem{key-23}\label{the:orlowski}Orlowski A, \textcolor{black}{\emph{Phys. Rev.
A}} \textbf{\textcolor{black}{48}} \textcolor{black}{(1993) 727.}
\bibitem{key-19}\textcolor{black}{\label{nonclassical}Dodonov V V, J. Opt. B. Quant.
and Semiclass. Opt. 4 (2002) R1.}
\bibitem{key-7}\textcolor{black}{\label{hbt}Hanbury-Brown R, Twiss R Q,} \textcolor{black}{\emph{Nature}}
\textbf{\textcolor{black}{177}} \textcolor{black}{27 (1956).}
\bibitem{key-21}\label{the:hilery2}Hillery M, \textcolor{black}{\emph{Phys. Rev.
A}} \textbf{\textcolor{black}{39}} \textcolor{black}{2994 (1989).}
\bibitem{key-13}\textcolor{black}{\label{hong1}Hong C K and Mandel L,} \textcolor{black}{\emph{Phys.
Rev. Lett.}} \textbf{\textcolor{black}{54}}
\textcolor{black}{(1985) 323.}
\bibitem{key-14}\textcolor{black}{\label{hong2}Hong C K and Mandel L,} \textcolor{black}{\emph{Phys.
Rev. A}} \textbf{\textcolor{black}{32}} \textcolor{black}{(1985)
974.}
\bibitem{key-8}\textcolor{black}{\label{lee1}Lee C T,} \textcolor{black}{\emph{Phys.
Rev. A}} \textbf{\textcolor{black}{41}} \textcolor{black}{1721
(1990).}
\bibitem{key-9}\textcolor{black}{\label{lee2}Lee C T,} \textcolor{black}{\emph{Phys.
Rev. A}} \textbf{\textcolor{black}{41}} \textcolor{black}{1569
(1990).}
\bibitem{key-16}\textcolor{black}{\label{hillery} Hillery M,} \textcolor{black}{\emph{Phys.
Rev. A}} \textbf{\textcolor{black}{36}} \textcolor{black}{3796
(1987).}
\bibitem{key-17}\textcolor{black}{\label{giri}Giri D K and Gupta P S} \textcolor{black}{\emph{Optics
Communication}} \textbf{\textcolor{black}{221}}
\textcolor{black}{(2003) 135.}
\bibitem{6}\textcolor{black}{\label{fernandez}Pathak A and Fernandez F M,} \textcolor{black}{\emph{Phys.
Lett. A}} \textbf{\textcolor{black}{341}} \textcolor{black}{(2005)
390.}
\bibitem{7}\textcolor{black}{\label{pathak1}Pathak A,} \textcolor{black}{\emph{Phys.
Lett. A}} \textbf{\textcolor{black}{317}} \textcolor{black}{(2003)
108.}
\bibitem{key-10}\textcolor{black}{\label{ba an}An N B,} \textcolor{black}{\emph{J.
Opt. B: Quantum Semiclass. Opt}}\textcolor{black}{.}
\textbf{\textcolor{black}{4}} \textcolor{black}{(2002) 222.}
\bibitem{key-20}\textcolor{black}{\label{the:martin}Pathak A and Garcia M, Communicated}\end{thebibliography}
\end{document}